\documentclass{aa} 

\input psfig.sty 
     
\begin{document}     
     
     
\title{An attempt to identify the extended synchrotron  
structure associated with the micro-quasar GRS 1915+105}  
  
\author{M. Ostrowski\inst{1} \and E. F\"urst\inst{2}} 
     
\institute{Obserwatorium Astronomiczne, Uniwersytet Jagiello\'nski, 
ul.Orla 171, 30-244 Krak\'ow, Poland \and 
Max-Planck-Institut f\"ur Radioastronomie, Auf dem 
H\"ugel 69, D-53-201 Bonn 1, Germany }   
     
\offprints{MO, e-mail: mio\@@oa.uj.edu.pl} 
     
\date{Received 18 October 2000; accepted 7 November 2000.} 
 
\titlerunning{An attempt to identify the extended synchrotron structure 
of GRS 1915+105} 
 
\authorrunning{M. Ostrowski \& E. F\"urst} 
     
\abstract{ 
The energy ejected from the galaxy micro-quasar GRS 1915+105 in the form 
of jets is expected to lead to formation of an extended double 
lobe/hot-spot structure with an energy content comparable to an average 
supernova remnant. We used the Effelsberg 100m telescope at 10.45~GHz in 
attempt to identify such structures. For this distant galactic plane 
source no definite identification was possible, due to high 
confusion by numerous background sources; however, a few suspect 
structures were identified. 
     } 
 
\maketitle

\section{Introduction}     
  
Among seven recently known galactic sources of relativistic jets (cf. 
Mirabel \& Rodr\'{\i}guez 1999), the object GRS 1915+105~, a first 
discovered galactic micro-quasar, is in some respects the most 
interesting. From the X-ray binary harbouring a black hole candidate, 
jets are occasionally ejected with a velocity $V_j \approx 0.92 c$, as 
measured with the VLA (Mirabel \& Rodr\'{\i}guez 1994), or even reaching 
$\approx 0.98 c$, as derived from MERLIN observations at smaller spatial 
scales (Fender et al. 1999). In several years of observations a 
recurrent activity of the source was monitored (cf. Fender \& Pooley 
1998, Rodr\'{\i}guez \& Mirabel 1999, Yadav et al. 1999). 
  
With the estimated mass of ejected material of $2\cdot 10^{25}$~g, Mirabel 
\& Rodr\'{\i}guez (1994) derived a large value for the kinetic energy in the 
ejected components' bulk motion, $E_k = 3\cdot 10^{46}$ erg. They noted 
that the power of the process ejecting blobs of radio emitting plasma is 
$\sim 400$ times greater than the steady X-ray flux radiated at the same 
time. A detailed discussion of the energy output in the form of jets 
from this source (cf. Levinson \& Blandford 1996, Ghisellini 1999, 
Gliozzi et al. 1999, Rodr\'{\i}guez \& Mirabel 1999) shows that varying 
assumptions about ejected particles and the jet Poynting flux can 
substantially scale down this energy estimate for a single ejected blob, 
but the average kinetic power injected into the medium in the form of 
jets can be larger than the maximum steady photon luminosity of  $\sim 3 
\cdot 10^{38}$ erg/s. Additionally, Fender \& Pooley (2000) suggest that 
GRS 1915+105 can inject more energy and matter into the outflow during 
periods of repeated small events than it does during the large 
ejections. Thus, the source activity, expected to last at least several 
$10^5$ years, will pump into the surrounding medium energy equal to at least 
$\sim 10^{51}$ ergs, an amount comparable to the energy injected by an 
average supernova event. It is expected that processes dissipating this 
energy -- shock waves and the generation of turbulent plasma motions -- 
generate also populations of energetic electrons, which in turn may be 
observed through their synchrotron radiation. As a final result, formation 
of a large scale radio source with a classical double structure is 
expected, presumably similar to extragalactic FRII sources. As estimated 
by Levinson \& Blandford (1996) the source can reach a spatial extension 
of up to a few hundred parsecs. For the estimated GRS 1915+105 distance 
near $12$ kpc and the angle of jet ejections with respect to the line of 
sight close to $70^\circ$, the angular extension of the respective 
extended source may reach $\sim 1^\circ$ on the sky. 
  
A few objects among strong binary X-ray sources were discovered as 
radio stars, with small scale radio jets (cf. Hjellming 1988, Mirabel \& 
Rodr\'{\i}guez 1999). Among them, only two exhibit extended diffuse 
structures. The relativistic jets source SS433, situated two degrees 
below the Galactic plane, is influencing (powering) a $1^\circ \, \times 
\, 2^\circ$ synchrotron nebula, which is assumed to be a 
supernova remnant. The second, Circinus X-1, is at the Galactic 
plane and is surrounded by the $20$ arc-minute synchrotron source, 
elongated along the small-scale jet direction. The source Scorpius X-1 
is accompanied by two radio hot-spots at distances $\approx 1^\prime$. 
However, all these sources are much (a few orders of magnitude) weaker 
in X-rays and in their jet kinetic power than GRS 1915+105. Looking for 
analogies for GRS 1915+105, one could consider Cygnus X-3, a possible 
black hole candidate. In this source, relativistic ejections were 
observed at mili-arc-second scales, but there is no definite 
identification of a large scale structure associated with this object. 
  
Rodr\'{\i}guez \& Mirabel (1998, $\equiv$ RM98) performed a search of 
extended counterparts of GRS 1915+105 with the use of VLA at 20 cm 
(configuration D) to look for possible hot spots at jets' terminal 
shocks. They discovered bright spots coincident with infra-red sources 
IRAS 19132+1035 and IRAS 19124+1106. Detailed observations of these 
sources at a few VLA frequencies (20 cm, 6 cm and 2 cm) revealed 
a non-thermal feature extended along the jet position angle. Both 
sources were also observed with the use of H92$\alpha$ recombination 
line. The results are fully compatible with HII regions in distances 6 
and 7 kpc, and the mentioned non-thermal feature could be an 
unconfirmed indication of association with GRS 1915+105. Also, 
a continuation of this work (Chaty et al. 2000) does not lead to a clear 
conclusion in this matter. 
  
Independent from RM98, we performed a search of the above mentioned 
extended structure using the Effelsberg 100 m telescope at 10.45~GHz. As 
this instrument is more able to recognize diffuse synchrotron features, 
one can consider our observations as complimentary to the ones of RM98. 
Also, we considered a substantially larger region of $2^\circ \times 
2^\circ$ near GRS 1915+105. The performed observations did not provide 
any direct proof for the association of the observed synchrotron 
features with jets. With respect to RM98, two new more distant hot 
spots, coincident with the jet direction, are noted. 
  
\section{Observations}  
  
To plan new observations we analysed a $2^\circ \times 2^\circ$ region 
centred approximately at the position of GRS 1915+105 
($\alpha(1950)~=~19h~12m~49,966s, \delta(1950)~=~10^\circ~51'~26,73"$), 
based on existing data. Fig.~1 shows the superposition of the radio 
emission at an 11~cm wavelength, taken from the Effelsberg Galactic plane 
survey (Reich et al. 1984) on the IRAS 60~$\mu$m emission (Beichmann et 
al. 1985). Two complexes of HII regions dominate by their strong 
emission 
($\alpha(1950)~=~19h~11m~47s$, $\delta(1950)~=~11^\circ~07'~03"$, 
$\alpha(1950)~=~19h~11m~06s$, $\delta(1950)~=~10^\circ~48'~25"$, see 
Wink et al. 1982). The shell-type emission feature near 
$\alpha(1950)~=~19h~14m$, $\delta(1950)~=~11^\circ~4'$ was identified as 
a supernova remnant by F\"{u}rst et al. (1987). Within a cone of jet 
position angles measured by various authors (cf. Fender et al. 1999, 
Rodr\'{\i}guez \& Mirabel 1999, Sams et al. 1996) the one 
(PA~=~155$^\circ$), accurately determined by Mirabel \& Rodr\'{\i}guez 
(1994), is indicated in the figure. 
  
\begin{figure}                    
\vspace{8.6cm} 
\includegraphics{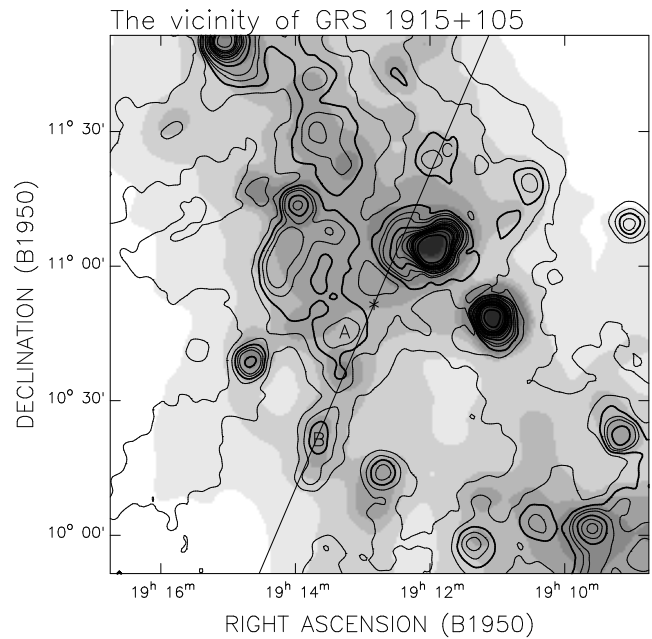} 
\caption[ ]{Two square degree vicinity of GRS 1915+105. An 11-cm Bonn  
map (contours) is superposed on the IRAS infrared map at $60$ $\mu$m  
(shaded intensity scale). The position of GRS 1915+105 is indicated with 
an asterisk, and the Mirabel \& Rodr\'{\i}guez (1994) jets' position with a 
solid line. Three regions observed by us are denoted with the 
symbols A, B, and C, respectively. } 
\end{figure}  
     
Along this line, three patchy emission features are visible, which have 
not yet been identified. They are denoted as A, B, and C. We have conducted 
new observations of these three regions  at 10.45~GHz with the 
Effelsberg 100-m-radiotelescope to identify the nature of the radio 
emission. We made use of the high-sensitivity four-feed receiver system to 
map the objects in total power and linear polarization. The observations 
were made by moving the telescope in azimuthal direction along the four 
feeds. The system uses the `software beam switching' technique (Morsi \& 
Reich 1988). The restoration procedure was introduced by Emerson et al. 
(1979). A detailed description of the system is given by Schmidt 
et al. (1993). Some parameters relevant for the current observation are 
summarized in Table~1. 
  
\medskip  
  
{\bf Table 1.} Parameters of the 10.45~GHz observations  
  
\begin{tabular}{ll} \hline   
Date of observation & 25 January 1998 \\ 
Frequency           & 10.45~GHz     \\  
Bandwidth           & 300~MHz       \\  
Calibrator          & 3C286         \\  
Calibrator flux density & 4.5~Jy    \\ Calibrator polarization & 11.7\%  
at 31$^\circ$ \\ HPBW                    & 69"                 \\  
T$_B$/S[K/Jy]           & 2.43                \\ Sensitivity I  
& 5.1~mKT$_B$       \\ Sensitivity PI          &  2.4~mKT$_B$      \\  
\hline \end{tabular}  
  
\medskip  
  
The results for total power of the three regions are combined on one map 
and displayed in Fig.~2. In polarization, no signal beyond 3~r.m.s was 
detected.

\section{A review of the observed features}  
  
All three objects found in the 11~cm data (Fig.~1) are also visible at 
10.45~GHz. We used the method of differential spectral index plots 
(TT-plots, see Turtle et al. 1962) to obtain the temperature spectral 
index $\beta_T$ ($T_B~\propto~\nu^{\beta_T}$) from the 11~cm survey 
data and the 10.45~GHz observations. For the three objects, we obtained 
the following spectral indices: 
  
\smallskip  
  
\begin{tabular}{lrr}  
Source A: & $\beta_T$~=~-2.41 & $\pm$0.2 \\  
Source B: & $\beta_T$~=~-2.09 & $\pm$0.1 \\  
Source C: & $\beta_T$~=~-2.03 & $\pm$0.1 \\  
\end{tabular}  
  
\medskip   
  
Source A is a faint extended-emission plateau which could be of 
nonthermal nature. The integrated flux density is about 1.1~Jy. At the 
bottom of this area, a compact source of 99~mJy is visible. It coincides 
with the infrared source IRAS~19132+1035 discussed by RM98 and Chaty et 
al. (2000).

\begin{figure}                    
\vspace{13.3cm} 
\includegraphics{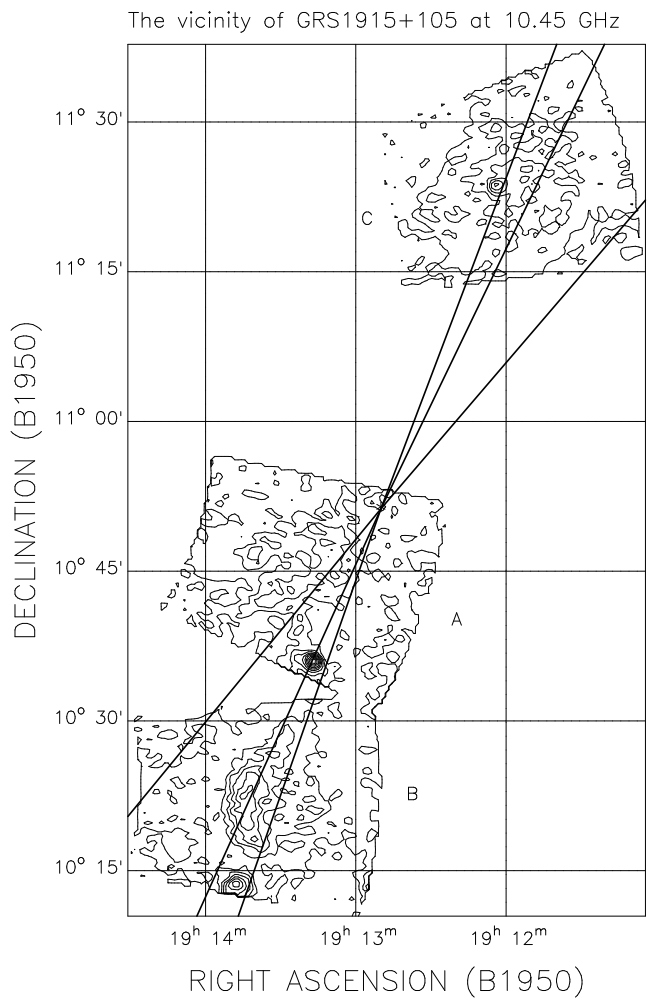} 
\caption[ ]{The 10.45~GHz Effelsberg maps of regions A, B and C from  
Fig.~1. The jet position lines given in the literature are 
plotted.}  
\end{figure}  
     
The sources in regions B and C, outside the range observed by RM98, are 
very likely thermal emission features, probably faint HII emission. As 
seen on Fig.~1, the region B contains an extended `jet-like' feature, 
partly with a substantial thermal component. The present 10.45~GHz 
observations reveal a shell-like morphology of this feature, 
characterized by a thermal-like spectrum. If this feature is produced 
by a jet-cloud interaction, the spectrum could be more complicated. In 
order to check this, additional observations at intermediate frequencies 
are required. The `hot-spot' at the bottom of Fig.~2 is characterized 
by a steeper non-thermal spectrum. 
  
A map of region C at the top of Fig.~2 shows an extended source with a 
thermal-like spectrum. It is characterized by a complicated morphology 
with a number of brighter spots over the structure. We note that the 
brightest of these knots (visible also at the edge of the RM98 VLA map) 
is situated close to the axis formed by GRS 1915+105, the hot spot of 
region A (coincident with the IRAS 19132+1035, cf. RM98) and the 
previously discussed hot-spot in region B. The second IRAS 19124+1106 
source studied by RM98 is situated outside fields observed by us, but 
also lies close to this axis. 
  
\section{Conclusions}  
     
Observations of a synchrotron nebula associated with a galactic source 
of relativistic jets would provide an opportunity to investigate 
processes of jet kinetic energy dissipation with much larger accuracy 
than in extragalactic sources. In particular, we are interested 
in observing the expected relativistic shock waves in jets' 
terminal points and/or interaction of such jets with molecular medium. 
Unfortunately, the position of GRS 1915+105 far away in the galactic 
plane, makes such observations difficult, due to effects of radiation 
attenuation, substantial depolarization effects at radio frequencies 
and superposition of many background/foreground sources 
occurring in the investigated region of the sky. 
  
Our Effelsberg 10.45~GHz observations performed within the $2^\circ 
\times 2^\circ$ region surrounding GRS 1915+105 reveal a number of 
structures situated along the jets' direction, but -- besides the 
positional coincidences -- without any clear relation to GRS 1915+105. 
We note three `hot-spots' (plus the RM98 source IRAS 19124+1106) 
situated along the same axis of the position angle, approximately 
$160^\circ$. All three are the brightest points at the respective maps. 
The object coincident with the IRAS source discussed by RM98 shows a 
non-thermal extension along the jet direction. A possible identification 
of relativistic shocks at the mentioned hot spot positions could be done 
if substantial high energy photon fluxes are detected from these 
places. A preliminary review of existing X-ray observations did not 
reveal any such sources (M. Ba{\l}uci\'{n}ska-Church \& M. Church -- private 
communication).

\begin{acknowledgements}     
  
We are grateful to M. Urbanik and M. Soida for their help with 
preliminary reduction of the data and to M. Ba{\l}uci\'{n}ska-Church \& M. 
Church  for review of the existing X-ray data and comments on the text. 
MO acknowledges support from the {\it Komitet Bada\'n Naukowych} through 
the grants PB 179/P03/96/11 and  PB 258/P03/99/17. 
  
\end{acknowledgements}

{}  
  
\end{document}